\documentclass[twocolumn,floatfix,showpacs,showkeys,preprintnumbers,nofootinbib,superscriptaddress]{revtex4}
\usepackage[utf8]{inputenc}
\usepackage[sort&compress]{natbib}
\usepackage{ulem}
\usepackage{bm}
\usepackage{times}
\usepackage{amssymb,amsbsy,amsmath,amsfonts}
\usepackage{graphicx}
\usepackage{float}
\usepackage{color}
\usepackage{morefloats}
\usepackage{rotating}
\usepackage{srcltx}
\usepackage{slashed}
\usepackage{subfigure}
\usepackage{multirow}
\usepackage{verbatim}
\usepackage{hyperref}
\usepackage{tabularx}
\usepackage{adjustbox}

\usepackage{overpic}
\usepackage{makecell}

\begin{document}

\title{Productions of $X(3872)$, $Z_c(3900)$, $X_2(4013)$, and $Z_c(4020)$ in $B_{(s)}$  decays offer strong clues on their molecular nature}

\author{Qi Wu}
\affiliation{Institute of Particle and Nuclear Physics, Henan Normal University, Xinxiang 453007, China}
\affiliation{School of Physics and Center of High Energy Physics,
Peking University, Beijing 100871, China}

\author{Ming-Zhu Liu}~\email{Corresponding author: zhengmz11@buaa.edu.cn}
\affiliation{
Frontiers Science Center for Rare isotopes, Lanzhou University,
Lanzhou 730000, China}
\affiliation{School of Physics, Beihang University, Beijing 102206, China}

\author{Li-Sheng Geng}~\email{Corresponding author: lisheng.geng@buaa.edu.cn}

\affiliation{School of Physics, Beihang University, Beijing 102206, China}
\affiliation{Peng Huanwu Collaborative Center for Research and Education, Beihang University, Beijing 100191, China}
\affiliation{Beijing Key Laboratory of Advanced Nuclear Materials and Physics, Beihang University, Beijing, 102206, China}
\affiliation{Southern Center for Nuclear-Science Theory (SCNT), Institute of Modern Physics, Chinese Academy of Sciences, Huizhou 516000, China}

\date{\today}
\begin{abstract}
 The exotic states $X(3872)$  and $Z_c(3900)$ have long been conjectured as isoscalar and isovector $\bar{D}^*D$ molecules. 
 In this work, we first propose the triangle diagram mechanism to investigate their productions in $B$   decays as well as their heavy quark spin symmetry partners, $X_2(4013)$ and $Z_c(4020)$.  We show that the large isospin breaking of the ratio $\mathcal{B}[B^+ \to X(3872) K^+]/\mathcal{B}[B^0 \to X(3872) K^0] $ can be attributed to the isospin breaking of the neutral and charged $\bar{D}^*D$  components in their wave functions.  For the same reason, the branching fractions of $Z_c(3900)$ in $B$ decays are smaller than the corresponding ones of $X(3872)$ by at least one order of magnitude, which naturally explains its non-observation. A hierarchy for the production fractions of $X(3872)$, $Z_c(3900)$, $X_2(4013)$, and $Z_c(4020)$ in $B$ decays, consistent with all existing data, is predicted.  
  Furthermore,  with the factorization ansatz we extract the decay constants of $X(3872)$, $Z_c(3900)$,   and $Z_c(4020)$ as   $\bar{D}^*D^{(*)}$ molecules via the $B$  decays, and then calculate their branching fractions in the relevant $B_{(s)}$ decays, which turn out to agree with all existing experimental data.      The mechanism we proposed is useful to elucidate the internal structure of the many exotic hadrons discovered so far and to extract the decay constants of hadronic molecules, which can be used to predict their production in related processes. 
\end{abstract}


\maketitle

\section{Introduction}

 The $X(3872)$ was discovered in 2003 by the Belle Collaboration ~\cite{Belle:2003nnu} and later confirmed in many other experiments~\cite{BaBar:2004iez,CDF:2003cab,D0:2004zmu,CMS:2013fpt,LHCb:2011zzp,BESIII:2013fnz,LHCb:2013kgk}. 
Its mass, $3871.69\pm0.17$ MeV,  is lower than the prediction of the legendary Goldfrey-Isgur quark model~\cite{Godfrey:1985xj} by almost 80~MeV.   In addition, the ratio $\mathcal{B}[X(3872)\to J/\psi \pi^+\pi^-\pi^0]/\mathcal{B}[X(3872)\to J/\psi\pi^+\pi^-]$~\cite{Belle:2005lfc,BaBar:2010wfc,BESIII:2019qvy}  shows large isospin-breaking effects, difficult to understand for a conventional charmonium. 
In 2013, the BESIII Collaboration and Belle Collaboration observed a charged charmonium-like state $Z_{c}(3900)$ in the $J/\psi \pi^{\pm}$ mass distribution of $e^{+}e^{-}\to J/\psi \pi^{+}\pi^{-}$~\cite{BESIII:2013ris,Belle:2013yex}, which is above the mass threshold of $\bar{D}^*D$ and has naturally been explained as a $\bar{D}^*D$ resonant state and the isospin partner of $X(3872)$~\cite{Albaladejo:2015lob,Karliner:2015ina}. 
Treating $X(3872)$ and $Z_c(3900)$ as $\bar{D}^*D$ molecules~\cite{Swanson:2003tb,Voloshin:2003nt,AlFiky:2005jd,Liu:2008fh,Sun:2011uh,Nieves:2012tt,Guo:2013sya,Karliner:2015ina, Braaten:2005ai,Gamermann:2009fv,Ortega:2009hj,Hanhart:2011tn,Li:2012cs,Wang:2013daa,Takeuchi:2014rsa,Zhou:2017txt,Mutuk:2018zxs,Wu:2021udi,Wang:2020dgr}, heavy quark spin symmetry (HQSS) implies the existence of two  $\bar{D}^*D^*$ molecules,  a $J^{PC}=2^{++}$  bound state~\cite{Nieves:2012tt,Guo:2013sya,Baru:2016iwj} and a $J^{PC}=1^{+-}$ resonant state~\cite{Wang:2014gwa,Yang:2020nrt,Meng:2020ihj,Baru:2021ddn,Yan:2021tcp,Du:2022jjv}. The former may correspond to the  $X(4014)$ state recently discovered in the $\gamma\psi(2S)$ mass distribution of $\gamma\gamma\to \gamma\psi(2S)$  by the Belle Collaboration~\cite{Belle:2021nuv}, 
and the latter may correspond to the resonant state $Z_c(4020)$  discovered in the $\pi^{\pm}h_{c}$ mass distribution of $e^{+}e^{-}\to h_{c} \pi^{+}\pi^{-}$ by the BESIII Collaboration~\cite{BESIII:2013ouc}.  
 Unlike their masses and decay patterns, their productions (particularly in $B_{(s)}$ decays) remain largely unexplored.     It is the purpose of the present work to fill this gap and show how the $\bar{D}^{(*)}D^*$ molecular picture explains simultaneously their productions, masses as well as decays, and thus help pin down their molecular nature in a highly nontrivial way.

The production mechanism of $X(3872)$ in $B$ decays was first proposed by  Braaten et al.~\cite{Braaten:2004fk,Braaten:2004ai}, where the $B$ meson first decays into $\bar{D}^{\ast}DK$ and then the charmed mesons rescatter and dynamically generate the $X(3872)$.  The predicted ratio $\mathcal{B}[B^0\to X(3872)K^0]/\mathcal{B}[B^+\to X(3872)K^+]$ depends on two unknown parameters and the resulting natural value for the ratio is one order of magnitude smaller than its experimental counterpart. We note that this ratio is reasonably described in Ref.~\cite{Wang:2022xga}, but not the absolute branching fractions.  
It is important to note that up to now, a complete understanding of the ratio $\mathcal{B}[B^0\to X(3872)K^0]/\mathcal{B}[B^+\to X(3872)K^+]$ and the absolute branching fractions in a unified framework is still missing.  In addition,   $X(3872)$ was  observed in other  decays  such as  $B^+ \to X(3872)K^{\ast+}$ and  $B_s^0 \to X(3872) \phi$~\cite{ParticleDataGroup:2022pth}. It is of vital importance to understand the branching fractions of $X(3872)$ as a $\bar{D}^*D$ molecule in the $B_{(s)}$ decays.   Another puzzle related to the four $\bar{D}^*D^{(*)}$ molecules is that although the $X(3872)$ has been observed in multiple channels of $B_{(s)}$ decays, the other three have not, which calls for an explanation. Furthermore, for planning future experiments, it is imperative to know the branching fractions of the other three states in $B_{(s)}$ decays,   given the fact that $B_{(s)}$ decays have served as important discovery channels for many exotic states and more data can be expected in near future~\cite{Belle-II:2018jsg}. 


\section{Theoretical framework}
\label{theoretical}

\begin{figure*}[htp]
\centering
\includegraphics[width=1.75\columnwidth]{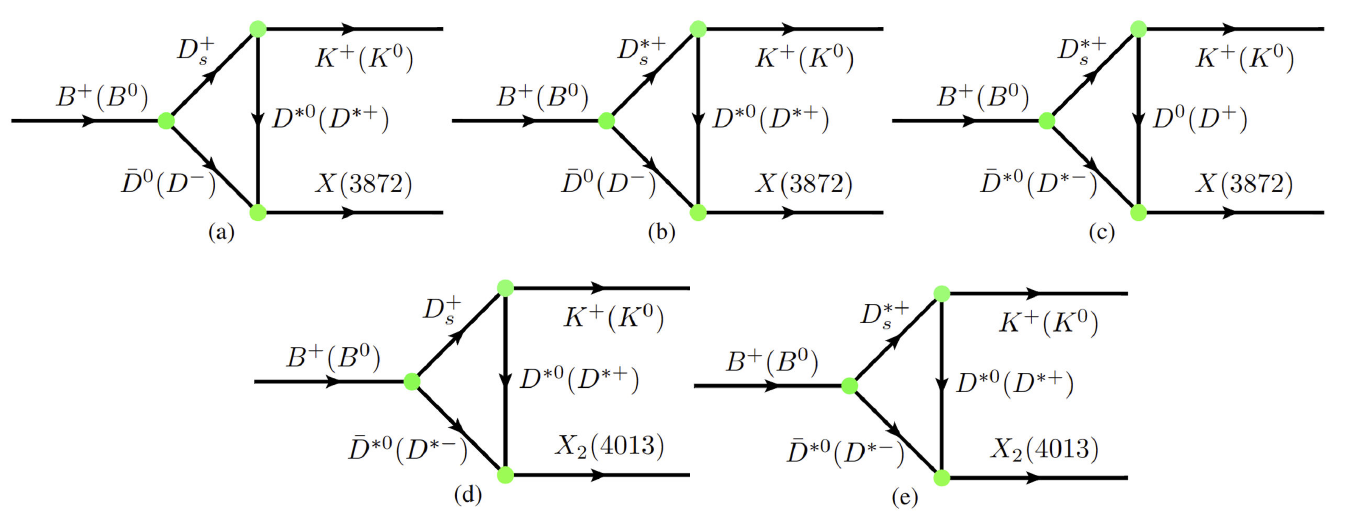}
\caption{\label{3872tri}
Triangle diagrams accounting for (a-c) $B^{+}({B}^0) \to D_s^{(*)+}\bar{D}^{(*)0}(D^{(\ast)+})\to X(3872)K^{+}(K^0)$ and (d-e) $B^{+}({B}^{0}) \to D_s^{(*)+}\bar{D}^{(*)0}(D^{(\ast)+})\to X_2(4013)K^+(K^0)$.}
\end{figure*}

 In this work, we propose the triangle mechanism to account for the productions of the  $X(3872)$, $Z_c(3900)$, $X_2(4013)$, and $Z_c(4020)$ molecules in $B$  decays. In this mechanism,  the $B$ meson first weakly decays into a pair of charmed mesons $D_{s}^{(\ast)}\bar{D}^{(\ast)}$, which proceeds via the external $W$-emission mechanism at the quark level as shown in Fig.~\ref{quark}~(a) in Appendix~\ref{appenda}. We only consider the external $W$-emission mechanism because it is usually the dominant one~\cite{Chau:1982da,Chau:1987tk,Molina:2019udw}. As shown later, our results corroborate this assumption.    Next, the charmed-strange mesons $D_{s}^{(\ast)}$ decay into a charmed meson $D^{(\ast)}$ and a kaon.  Finally the $\bar{D}D^*$ and $\bar{D}^*D^*$ molecules are dynamically generated via the final-state interactions of $\bar{D}^{(*)}D^*$ as shown in Fig.~\ref{3872tri}.   Here the isoscalar  $\bar{D}D^*$ and $\bar{D}^*D^*$ molecules refer to $X(3872)$ and $X_{2}(4013)$,  and their isovector counterparts are   $Z_c(3900)$ and $Z_c(4020)$.   We do not explicitly present the triangle diagrams for the $Z_{c}(3900)$ and $Z_{c}(4020)$, which can be obtained by replacing the $X(3872)$ and  $X_2(4013)$ of Fig.~\ref{3872tri} with $Z_c(3900)^0$ and $Z_c(4020)^0$, respectively.

We note in passing that the triangle mechanism has been applied to study the productions of $D_{s0}^*(2317)$, $D_{s1}(2460)$~\cite{Liu:2022dmm}, $D_s^+D_s^-$, and $D\bar{D}$ molecules~\cite{Xie:2022lyw}, yielding branching fractions in agreement with data. However, in the present work, because of the existence of a complete multiplet of hadronic molecules and of the interplay between the charged and neutral components in the wave functions of these states, there is richer physics, such as the isospin-breaking ratios and the nontrivial hierarchy among the branching ratios. As a result, the productions studied are more informative and play a more decisive role in disclosing the nature of $X(3872)$, $Z_c(3900)$, and their HQSS partners, $X_2(4013)$ and $Z_c(4020)$.


We employ the effective Lagrangian approach to calculate the Feynman diagrams of Fig.~\ref{3872tri}.  
The relevant Lagrangians describing the interactions of each vertex in the triangle diagrams  and the determination of the corresponding couplings either by fitting to data or relying on symmetries are presented in Appendix~\ref{appenda}, Appendix~\ref{appendb}, and Appendix~\ref{appendc}.  
It is straightforward to calculate the Feynman diagrams of Fig.~\ref{3872tri} and obtain the following amplitudes 

\begin{widetext}
\begin{eqnarray}  
\label{3872amp}
\mathcal{A}_{a} & = &\int \frac{d^{4} q_{3}}{(2 \pi)^{4}} \frac{{\rm i}\mathcal{A}(B^{+}\to D_{s}^{+}\bar{D}^{0})\mathcal{A}\left(D_{s}^{+} \to {D}^{\ast 0 }K^{+}\right)\mathcal{A}\left({D}^{\ast 0 } \bar{D}^{ 0 } \to X(3872)\right)}{\left(q_{1}^{2}-m_{D_{s}^{+}}^{2}\right)\left(q_{2}^{2}-m_{\bar{D}^{0}}^{2}\right)\left(q_{3}^{2}-m_{{D}^{\ast0}}^{2}\right)},  \label{38721} \\
\mathcal{A}_{b} & =& \int \frac{d^{4} q_{3}}{(2 \pi)^{4}} \frac{{\rm i}\mathcal{A}(B^{+}\to D_{s}^{\ast+}\bar{D}^{0})\mathcal{A}\left(D_{s}^{\ast+} \to {D}^{\ast 0 }K^{+}\right)\mathcal{A}\left( {D}^{\ast 0 } \bar{D}^{ 0 } \to X(3872)\right)}{\left(q_{1}^{2}-m_{D_{s}^{\ast+}}^{2}\right)\left(q_{2}^{2}-m_{\bar{D}^{0}}^{2}\right)\left(q_{3}^{2}-m_{{D}^{\ast0}}^{2}\right)},
 \label{38722} \\
\mathcal{A}_{c} & =& \int \frac{d^{4} q_{3}}{(2 \pi)^{4}} \frac{{\rm i}\mathcal{A}(B^{+}\to D_{s}^{\ast+}\bar{D}^{\ast0})\mathcal{A}\left(D_{s}^{\ast+} \to {D}^{ 0 }K^{+}\right)\mathcal{A}\left( {D}^{ 0 } \bar{D}^{ \ast 0 } \to X(3872)\right)}{\left(q_{1}^{2}-m_{D_{s}^{\ast+}}^{2}\right)\left(q_{2}^{2}-m_{\bar{D}^{\ast0}}^{2}\right)\left(q_{3}^{2}-m_{{D}^{0}}^{2}\right)},   \label{38723} \\
\mathcal{A}_{d} & =& \int \frac{d^{4} q_{3}}{(2 \pi)^{4}} \frac{{\rm i}\mathcal{A}(B^{+}\to D_{s}^{+}\bar{D}^{\ast0})\mathcal{A}\left(D_{s}^{+} \to {D}^{\ast 0 }K^{+}\right)\mathcal{A}\left({D}^{\ast 0 } \bar{D}^{ \ast 0 } \to X_2(4013)\right)}{\left(q_{1}^{2}-m_{D_{s}^{+}}^{2}\right)\left(q_{2}^{2}-m_{\bar{D}^{\ast0}}^{2}\right)\left(q_{3}^{2}-m_{{D}^{\ast0}}^{2}\right)},  \\
\mathcal{A}_{e} & =& \int \frac{d^{4} q_{3}}{(2 \pi)^{4}} \frac{{\rm i}\mathcal{A}(B^{+}\to D_{s}^{\ast+}\bar{D}^{\ast0})\mathcal{A}\left(D_{s}^{\ast+} \to {D}^{\ast 0 }K^{+}\right)\mathcal{A}\left( {D}^{\ast 0 } \bar{D}^{\ast 0 } \to X_2(4013)\right)}{\left(q_{1}^{2}-m_{D_{s}^{\ast+}}^{2}\right)\left(q_{2}^{2}-m_{\bar{D}^{\ast0}}^{2}\right)\left(q_{3}^{2}-m_{{D}^{\ast0}}^{2}\right)},
\end{eqnarray}
\end{widetext}
where  $q_{1}$, $q_{2}$, and $q_3$  denote the momenta of   $D_{s}^{(\ast)}$, $\bar{D}^{(\ast)}$, and $D^{(\ast)}$, and the amplitudes for each vertex of the triangle diagrams are listed in the Supplemental Material.

\begin{figure*}[htbp]
\centering
\includegraphics[width=1.60\columnwidth]{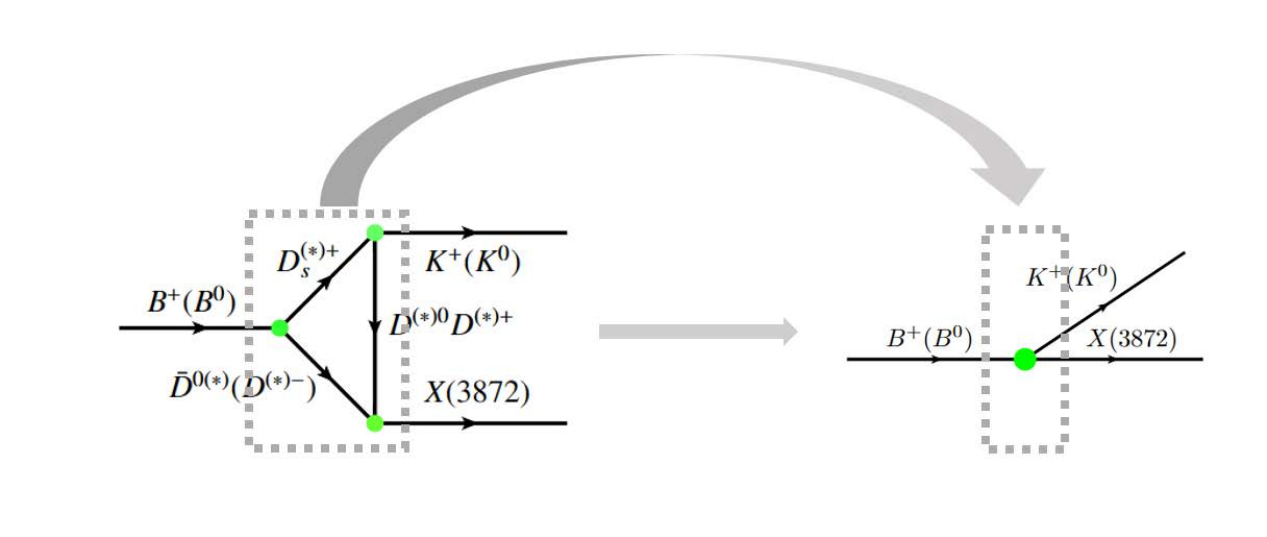}
\caption{\label{rtis}
Triangle diagrams illustrating  the decays $B^{+}(B^{0})\to K^{+}(K^0)X(3872)$ simplified as  tree diagrams.  
}
\end{figure*}

As shown in Fig.~\ref{rtis}, one can condense the triangle diagram into one vertex, leading to an effective description of the weak decay $B\to X(3872)K$ at the tree level. 
With the factorization ansatz,  the decay $B \to X(3872) K$  actually can be expressed as the product of two matrix elements:
\begin{widetext}
\begin{eqnarray}\label{BtoX387245}
\mathcal{A}\left(B \to X(3872) K\right)&=&\frac{G_{F}}{\sqrt{2}} V_{cb}V_{cs} a_{2}\left\langle X(3872)|(c\bar{c})| 0\right\rangle\left\langle K |(s \bar{b})| B\right\rangle,
\end{eqnarray}
\end{widetext}
where the effective Wilson coefficient   $a_2$  is   determined by reproducing the branching fractions of the   decay $B\to J/\psi K $  since the  $J/\psi$ can be viewed as a pure $c\bar{c}$.  The matrix element $\left\langle K |(s \bar{b})| B\right\rangle$  is characterized by form factors and the other one is expressed as $\left\langle X(3872)|(c\bar{c})| 0\right\rangle= m_{X(3872)} f_{X(3872)} \varepsilon^{\mu} $, where the decay constant  $f_{X(3872)}$ is unknown.        Using the equivalence of the triangle diagrams and tree diagrams in Fig.~\ref{rtis},   we can extract the decay constant of $X(3872)$ as a $\bar{D}^*D$ molecule, which is different from the estimation of the  $X(3872)$ decay constant as an excited charmonium state~\cite{Wang:2007sxa}. One can see that the molecular information of $X(3872)$ is well hidden in the $X(3872)$ decay constant.    Since only the tensor current for matrix element  $\langle X_{2}(4013)|\bar{c}c| 0\rangle$ is allowed~\cite{Cheng:2010yd}, the corresponding current for the matrix element  $\left\langle K |(s \bar{b})| B\right\rangle$  must be tensor, which is difficult to calculate.  Therefore, we can not directly extract the decay constant of $X_{2}(4013)$ along this line and only focus on the other three in this work.  
Following the strategy outlined, we can extract the decay constants of  $Z_c(3900)$ and $Z_c(4020)$ as hadronic molecules, and the corresponding current matrix elements are written as  $\left\langle Z_c(3900)|(c\bar{c})| 0\right\rangle= m_{Z_c(3900)} f_{Z_c(3900)} \varepsilon^{\mu} $ and    $\left\langle Z_c(4020)|(c\bar{c})| 0\right\rangle= m_{Z_c(4020)} f_{Z_c(4020)} \varepsilon^{\mu} $.

Now that the decay constants of $\bar{D}^{(*)}D^{(*)}$ molecules are obtained, it is straightforward to calculate the production rates of $\bar{D}^{(*)}D^{(*)}$ molecules in other $B_{(s)}$  decays. Here we choose the decay $B_{s} \to X(3872) \phi$ as an example to demonstrate the procedure.  Using the naive factorization approach,  the amplitude of the decay $B_{s} \to X(3872) \phi$ is expressed as
\begin{widetext}
\begin{eqnarray}\label{BstoX3872}
\mathcal{A}\left(B_s \to X(3872) \phi\right)&=&\frac{G_{F}}{\sqrt{2}} V_{cb}V_{cs} a_{2}\left\langle X(3872)|(c\bar{c})| 0\right\rangle\left\langle\phi |(s \bar{b})| B_s\right\rangle,
\end{eqnarray} 
\end{widetext}
where the effective Wilson coefficient $a_2$ is determined by reproducing the branching fraction of the weak decay  $B_{s} \to J/\psi \phi$, and the current matrix element $\left\langle\phi |(s \bar{b})| B_s\right\rangle$ is expressed as several form factors, which have the same form as the $B\to D^*$ form factors.    
The current  matrix element  $\left\langle X(3872)|(c\bar{c})| 0\right\rangle$ is already obtained via  the  decays $B \to X(3872) K$. Similarly, we can obtain the amplitudes of the weak decays of $B \to J/\psi K^*$ and $B_{s} \to J/\psi \eta$:
\begin{widetext}
\begin{eqnarray}\label{BstoX38721}
\mathcal{A}\left(B_s \to X(3872) \eta\right)&=&\frac{G_{F}}{\sqrt{2}} V_{cb}V_{cs} a_{2}\left\langle X(3872)|(c\bar{c})| 0\right\rangle\left\langle\eta |(s \bar{b})| B_s\right\rangle,  \nonumber \\ \mathcal{A}\left(B \to X(3872) K^*\right)&=&\frac{G_{F}}{\sqrt{2}} V_{cb}V_{cs} a_{2}\left\langle X(3872)|(c\bar{c})| 0\right\rangle\left\langle K^* |(s \bar{b})| B_s\right\rangle.
\end{eqnarray} 
\end{widetext}

 With  the amplitudes  for the weak  decays   given above, 
 one can compute the corresponding partial decay widths 
 \begin{eqnarray}
\Gamma=\frac{1}{2J+1}\frac{1}{8\pi}\frac{|\vec{p}|}{m_{B_{(s)}}^2}{|\overline{M}|}^{2},
\end{eqnarray}
where $J$ is the total angular momentum of the initial $B_{(s)}$ meson, the overline indicates the sum over the polarization vectors of final states, and $|\vec{p}|$ is the momentum of either final state in the rest frame of the $B_{(s)}$ meson.

\section{Results and discussions}
\label{results}

The couplings of $X(3872)$/$Z_c(3900)$ and their HQSS partners $X_2(4013)$/$Z_c(4020)$ to their constituents   $\bar{D}D^*$ and $\bar{D}^*D^*$ can be estimated in the contact range effective field theory approach~(See Appendix~\ref{appendd} for details).   
 As a $\bar{D}^*D$ bound state, $X(3872)$ contains both a neutral component     $\bar{D}^{*0}D^{0}/\bar{D}^{0}D^{\ast0}$ and a charged component  $D^{*+}D^{-}/D^{+}D^{\ast-}$ in its wave function.
  The couplings to the neutral and charged components are found to be, $g_{n}=3.86$~GeV and $g_{c}=3.39$~GeV, which indicates that the neutral component plays a more important role than the charged component, consistent with the conclusions of Refs.~\cite{Ortega:2009hj,Gamermann:2009fv,Li:2012cs,Guo:2014hqa,Zhou:2017txt,Yamaguchi:2019vea}. \footnote{  Once the couplings of $g_{n}$ and $g_{c}$ are obtained, we estimate the proportions of the neutral and charged components as $96\%$ and $4\%$, consistent with Refs.~\cite{Yamaguchi:2019vea,Wang:2023ovj,Song:2023pdq}   }  
  Employing HQSS, we can obtain the potentials of the $\bar{D}^{*0}D^{\ast0}/D^{*+}D^{\ast-}$ system and predict the existence of a $J^{PC}=2^{++}$ bound state with a mass of $m=4013.03$~MeV, corresponding to $X_2(4013)$. 
  Similarly, the $X_2(4013)$ couplings to its neutral and charged components are estimated to be  $g_{n}^{\prime}=5.36$~GeV and    $g_{c}^{\prime}=4.86$~GeV. Because the $Z_{c}(3900)$ is located above the mass thresholds of the neutral and charged components of $\bar{D}^*D$ by about 10 MeV, isospin-breaking effects are expected to be small. Therefore, we deal with the $Z_c$ states in the isospin limit. By reproducing the mass and width of $Z_{c}(3900)$, we obtain the coupling $g_{Z_{c}(3900)\bar{D}D^*}=7.10$~GeV. The HQSS dictates  the existence of  a $\bar{D}^*D^*$ molecule with  M$=4028$ MeV and $\Gamma=26$~MeV, whose coupling is estimated to be $g_{Z_{c}(4020)\bar{D}^*D^*}=1.77$.  In Table~\ref{couplingsr}, we present the ratios of the couplings in particle basis to those in isospin basis. 
For the isoscalar states, the couplings to the charged component and those to the neutral component are of the same sign, but for the isovector states, they are of the opposite sign, which has an important impact on our understanding of the productions of these molecules in $B$ decays as shown below. It is important to note that Table~\ref{couplingsr} only tells the relative sign between the neutral and charged components, while the relative size will be determined by data for the isoscalar molecules but assumed to be the same for the isovector molecules as discussed below and in Appendix~\ref{appendd}.

\begin{table}[!h]
    \centering
    \caption{Ratios of the couplings in particle basis to the  couplings in isospin basis.}
    \begin{tabular}{|c|cccc|}
    \hline
     Molecules    &~~~~${D}^{\ast+}D^-$  &~~~~${D}^{+}D^{\ast-}$ &~~~~${D}^{\ast0}\bar{D}^{0}$   &~~~~${D}^{0} \bar{D}^{\ast0}$ 
         \\ \hline
        $X(3872) $   &~~~~ $1/2$   &~~$-1/2$ &~~~~ $1/2$  &~~$-1/2$ \\
        $Z_c(3900)$   &~~~~ $1/2$    &~~~~ $1/2$  &~~$-1/2$  &~~$-1/2$
         \\ \hline  Molecules     &~~~~${D}^{\ast+}D^{\ast-}$ &~~~~${D}^{\ast0}\bar{D}^{\ast0}$ & &
         \\ \hline
        $X_2(4013) $   &~~~~ $1/\sqrt{2}$  &~~~~ $1/\sqrt{2}$  & &  \\
        $Z_c(4020)$  &~~~~ $1/\sqrt{2}$  &~~ $-1/\sqrt{2}$   &  & 
\\
    \hline
    \end{tabular}
   \label{couplingsr}
\end{table}

We employ the effective Lagrangian approach to calculate the branching fractions of $\bar{D}^*D^{(*)}$ molecules in $B$  decays illustrated in  Fig.~\ref{3872tri}, where the dominant uncertainties originate from the couplings of the three vertices of the triangle diagrams.   For the weak interaction vertices, the experimental uncertainties of the branching fractions of $B\to \bar{D}^{(\ast)}D_{s}^{(\ast)}$ lead to about $10\%$ uncertainty for the effective Wilson coefficient  $a_{1}$~\cite{Xie:2022lyw}\footnote{ The uncertainties of the experimental branching fractions of the weak decays $B\to \bar{D}^{(\ast)}D_{s}^{(\ast)}$  are transferred to the uncertainties of the product of effective Wilson coefficients and decay constants.  Since the uncertainties of decay constants $f_{D_s}$ and $f_{D_s^*}$ are small according to lattice QCD~\cite{Chen:2020qma}, the experimental uncertainties are only embodied into the effective Wilson coefficients.     }
For the vertices describing the dynamical generation of hadronic molecules, the uncertainties are mainly from the cutoff $\Lambda$ of the form factor. If we increase the cutoff from 1  to $2$ GeV,  the couplings decrease by about $10\%$. Therefore, we assign a $10\%$ uncertainty for the couplings of the molecules to their constituents~\cite{Xie:2022lyw}, a bit larger than the estimation for a cutoff variation from $0.5$ GeV to $1$ GeV~\cite{Guo:2014hqa}.   As for  the couplings  $g_{D_{s}^{(\ast)}D^{(\ast)}K}$
the large SU(4)-flavor symmetry breaking can lead to an uncertainty of about $33\%$~\footnote{The SU(3)-flavor symmetry breaking can be characterized by the difference between the decay constants $f_{K}$  and $f_{\pi}$, which is about $19\%$~\cite{Follana:2007uv,Carrasco:2014poa,Miller:2020xhy}. Along this line, the SU(4)-flavor symmetry breaking is estimated to be about $33\%$ by comparing the decay constants $f_{D}$ and $f_{K}$~\cite{Follana:2007uv,Carrasco:2014poa}, consistent with Ref.~\cite{Fontoura:2017ujf}.}. Finally, we obtain the uncertainties of the branching fractions originating from the uncertainties of these parameters via a  Monte Carlo sampling in their 1$\sigma$ intervals. One should note that there exists no extra free parameter in our model.

\begin{figure}[htbp]
  \centering
 \includegraphics[width=9cm]{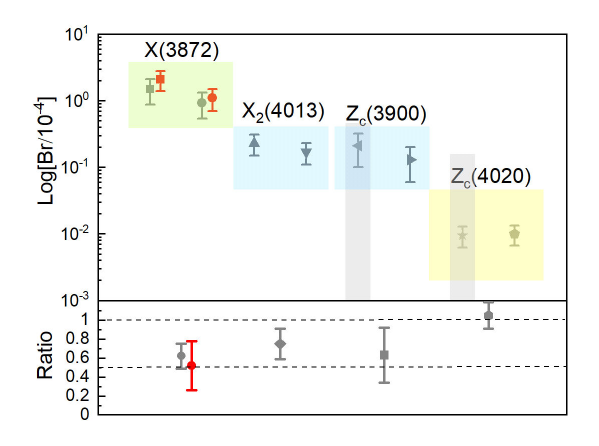}
  \caption{\label{Fig:BF} Top: branching fractions of $B^{+(0)}\to X(3872)K^{+(0)}$ (green block), $B^{+(0)}\to X_2(4013)K^{+(0)}$ (blue block), $B^{+(0)}\to Z_c(3900)K^{+(0)}$ (blue block), and $B^{+(0)}\to Z_c(4020)K^{+(0)}$ (yellow block). The left and right data points in each block are for the $B^+$ and $B^0$ decays, respectively. Bottom: the corresponding ratios between the branching fractions of $B^+$  and those of $B^0$. The red error bars and shadow parts are the corresponding experimental data.}
\end{figure}

In Fig.~\ref{Fig:BF}, we compare the predicted branching fractions of $\bar{D}^*D$ and $\bar{D}^*D^*$ molecules in $B$ decays with the available experimental data. The numbers are given in Table \ref{resultsd} of  Appendix~\ref{appende}. 
One can see that the branching fractions of the decays   $B^+\to X(3872)K^+$ and $B^0\to X(3872)K^0$ are in reasonable agreement with the experimental data. We further compute  the ratio  $\mathcal{B}[B^0\to X(3872)K^0]/\mathcal{B}[B^+\to X(3872)K^+]$ to be $0.62\pm 0.13$,  in agreement with the experimental value $0.52\pm0.26$ within uncertainties.  We note that the uncertainty of the predicted ratio is much smaller than that of the branching fractions. 
 We stress that the fact that the branching fractions of $X(3872)$ in $B$ decays can be reproduced in the $\bar{D}^*D$ molecular picture provides non-trivial support for the nature of $X(3872)$ as a $\bar{D}^*D$ bound state.

The branching fractions of  $\mathcal{B}[B^+\to Z_{c}(3900)K^+]$ and  $\mathcal{B}[B^0\to Z_{c}(3900)K^0]$ turn out to be $(1.0\sim 3.3)\times 10^{-5}$ and  $(0.6\sim 2.0)\times 10^{-5}$.  The upper limit of the experimental branching fraction $\mathcal{B}[B^+\to Z_{c}(3900)(Z_c(3900)\to \eta_{c}\pi^+\pi^-)K^+]$ is $4.7\times 10^{-5}$~\cite{ParticleDataGroup:2022pth}. Although due  to the unknown  branching fraction of $\mathcal{B}[Z_c(3900)\to \eta_{c}\pi^+\pi^-]$, we can not determine  $\mathcal{B}[B^+\to Z_{c}(3900)K^+]$, our prediction is safely below the experimental upper limit. We note that the ratio $\mathcal{B}[B^+\to Z_{c}(3900)K^+]/\mathcal{B}[B^0\to Z_{c}(3900)K^0]=0.63\pm0.29$ shows large isospin-breaking effects. However, unlike the case of $X(3872)$ and $X_2(4013)$, this is not due to isospin breaking of the wave functions but is mainly caused by the Wilson coefficient $a_1$ fitted to the $B^{+(0)}\to D_s^{+} \bar{D}^0( D^{-}) $ and $B^{+(0)}\to D_s^{\ast+} \bar{D}^{\ast0}( D^{\ast-}) $  decays (see Appendix~\ref{appende} for details).   
It is interesting to compare the branching fractions of   $\mathcal{B}[B\to Z_{c}(3900)K]$  with those of     $\mathcal{B}[B\to X(3872)K]$. The former is smaller than the latter by one order of magnitude, which is consistent with the fact that the $Z_{c}(3900)$ state has not been observed in $B$  decays.   We note that only the amplitude of  Fig.~\ref{3872tri}~(a)  and that  of Fig.~\ref{3872tri}~(c) 
contribute to the decays  of the $B$ meson into  the $\bar{D}^*D$ molecules, while the contribution of Fig.~\ref{3872tri}~(b) is accidentally very small. The sign of the amplitude of  Fig.~\ref{3872tri}~(a)  and that of Fig.~\ref{3872tri}~(c)  depend on the relative sign between the charged and neural components in the wave functions of the  $\bar{D}^*D$ molecules.  From Table~\ref{couplingsr} one can see that the sign is opposite for the isocalar molecules but the same for the isovector molecules. As the two amplitudes for the isoscalar molecules add constructively, but those for the isovector molecules add destructively, the production rates of $Z_{c}(3900)$ in $B$  decays are lower than those of $X(3872)$ in $B$  decays.

We now turn to the branching fractions of $X_2(4013)$ and $Z_c(4020)$ in $B$  decays.  The predicted branching fractions of  $\mathcal{B}[B^{+}\to X_2(4013) K^{+}]$ and  $\mathcal{B}[B^{0}\to X_2(4013) K^{0}]$ are $(1.5\sim 3.1)\times 10^{-5}$ and  $(1.1\sim 2.3)\times 10^{-5}$, and the ratio  $\mathcal{B}[B^{0}\to X_2(4013) K^{0}]/\mathcal{B}[B^{+}\to X_2(4013) K^{+}]$ is estimated to be $0.75\pm0.16$. 
We note that the isospin breaking of the ratio is mainly caused by the isospin breaking of the $\bar{D}^*D^*$ wave function.    Similarly, we predict the branching fractions  $\mathcal{B}[B\to Z_c(4020) K]$ to be around $1\times 10^{-6}$, which are lower than those  of $\mathcal{B}[B\to Z_c(3900) K]$ as well as  $\mathcal{B}[B\to X_2(4013) K]$  by one order of magnitude. This implies that it will
be more difficult to observe them in $B$ decays. 

 \begin{table}[ttt]
\centering
\caption{ Decay constants (in units of MeV) of  $X(3872)$, $Z_c(3900)$, and $Z_c(4020)$   as  $\bar{D}^*D^{(*)}$ molecules.  \label{decayconstants12}
}
\begin{tabular}{|c c | }
\hline
    Molecules   &   Decay Constants    
         \\ \hline ${X(3872)}$    &$182.22^{+34.62}_{-42.98}$  
         \\
   
${Z_c(3900)}$    & $68.85^{+16.14}_{-21.33}$ \\  ${Z_c(4020)}$  &    $15.69^{+2.52}_{-3.01} $       \\    
\hline
\end{tabular}
\end{table}

\begin{table*}[htp]
\centering
\caption{Branching fractions ($10^{-4}$) of the decays  $B_s^{0}\to X(3872) \phi$, $B\to X(3872) K^*$, and $B_s^{0}\to X(3872) \eta$.  \label{resultq2}
}
\begin{tabular}{|c c c| c c c |c c c c c c c  }
\hline
    Decay modes   &    Exp.~\cite{ParticleDataGroup:2022pth}  &  $a_2$    &    Decay modes    &~~~~ Our predictions  &~~~~ Exp.~\cite{ParticleDataGroup:2022pth}  
         \\ \hline $B^{+}\to J/\psi K^+ $    &$10.20\pm0.19$    & $0.271^{+0.002}_{-0.003}$   & ~~~$B^{+}\to X(3872) K^{+} $    &~~~~ $1.49^{+0.62}_{-0.62}$  &~~~~$2.1\pm 0.7$     
                 \\   
$B^{+}\to J/\psi K^{\ast+} $    &$14.3\pm0.8$    & $0.236^{+0.007}_{-0.007}$   & ~~~$B^{+}\to X(3872) K^{\ast+} $    &~~~~ $3.47^{+0.85}_{-0.85}$  &~~~~ $2.8\sim 6$     
                 \\
 $B_s^{0} \to J/\psi \phi $   & $10.4\pm 0.4$  &  $0.206^{+0.004}_{-0.004}$  &    $B_s^{0} \to X(3872)\phi $   &~~~~ $2.39^{+0.58}_{-0.58}$   &~~~~ $1.1\pm 0.4$            
 \\   $B_s^{0} \to J/\psi  \eta $   & $4.0\pm 0.7$  &    $0.212^{+0.018}_{-0.019}$    &  $B_s^{0} \to X(3872)\eta $    &~~~~ $0.41^{+0.11}_{-0.11}$  &~~~~ $-$      
         \\ 

\hline
\end{tabular}
\end{table*}

The production mechanism of the  $\bar{D}^*D^{(*)}$  molecules in $B$  decays via the triangle diagrams can be simplified as tree-level diagrams. This way, one can extract the decays constants of $X(3872)$, $Z_c(3900)$, and $Z_{c}(4020)$ as  $\bar{D}^*D^{(*)}$ molecules.  From Fig.~\ref{rtis}, we can see  that the summation of  Eq.(\ref{38721}),  Eq.(\ref{38722}) and Eq.(\ref{38723}) representing the amplitude of the triangle diagram  is equal to   Eq.(\ref{BtoX387245}) representing the amplitude of the tree-level diagram, where the former amplitude was already calculated but with 
 two unknown parameters $a_2$ and $f_{X(3872)}$ left for the latter amplitude.    
First, we determine the effective Wilson coefficient  $a_2=0.271^{+0.002}_{-0.003}$ by reproducing  the experimental  branching fraction $\mathcal{B}(B^+\to J/\psi K^+)$.  Then   we  extract     the   decay constant  of $X(3872)$ as a $\bar{D}^*D$ molecule, e.g.,  $f_{X(3872)}=182.22^{+34.62}_{-42.98}$~MeV.
 The decay constant of $X(3872)$ as a purely excited charmonium state is estimated to be  329 MeV~\cite{Wang:2007sxa} or 335 MeV~\cite{Liu:2007uj}, which is much larger than that as a hadronic molecule.  Once the $X(3872)$ decay constant is obtained~\footnote{The  $X(3872)$ can also contain a $c\bar{c}$ component~\cite{Matheus:2009vq,Zanetti:2011ju}. The studies performed in this work show that the $\bar{D}^*D$ component plays a dominant role.}, one can predict the branching fractions of these states in other processes, such as  $B \to K^*$, $B_s \to \eta$, and $B_s \to \phi$, which share the same production mechanism as that of $B \to X(3872) K$ at the quark level. The unknown parameters of the form factors of  these hadron  transitions   are taken from Table~\ref{BtoKformfactor1} in Appendix~\ref{appenda}, and the corresponding  effective Wilson coefficients $a_2$  are determined by the experimental branching fractions of the  decays $B \to 
 J/\psi K^*$, $B_s \to J/\psi\eta$, and $B_s \to J/\psi \phi$ listed  in Table~\ref{resultq2}. 
     With the $X(3872)$  decay constant determined, we can obtain the branching fractions:  $\mathcal{B}(B^+ \to X(3872) K^{*+})=3.47^{+0.85}_{-0.85}\times 10^{-4}$ and  $\mathcal{B}(B_s^0 \to X(3872) \phi)=2.39^{+0.58}_{-0.58}\times 10^{-4}$, consistent with the experimental data.   Similarly, we predict the branching fraction  $\mathcal{B}(B_s^0 \to X(3872) \eta)$ to be $0.41^{+0.11}_{-0.11}\times 10^{-4}$, which can be verified by  future experiments.

\begin{table*}[ttt]
\centering
\caption{Branching fractions ($10^{-6}$) of the decays $B_s^{0}\to Z_c(3900) \phi$, $B\to Z_c(3900) K^*$, $B_s^{0}\to Z_c(3900) \eta$, $B_s^{0}\to Z_c(4020) \phi$, $B\to Z_c(4020) K^*$, and $B_s^{0}\to Z_c(4020) \eta$.  \label{resultsprediction45}}
\begin{tabular}{|c c | c c  |c c c c c c c  }
\hline
    Decay modes   &    Our Predictions   &     Decay modes    &~~~~ Our predictions  
         \\ \hline 
$B^{+}\to Z_c(3900) K^{\ast+} $    &$49.64^{+15.15}_{-15.15}$     & ~~~$B^{+}\to Z_c(4020)K^{\ast+} $    &~~~~ $2.51^{+0.51}_{-0.51}$   
                 \\
 $B_s^{0} \to Z_c(3900) \phi $   & $34.07^{+10.36}_{-10.36}$   &    $B_s^{0} \to Z_c(4020)\phi $   &~~~~ $1.63^{+0.33}_{-0.33}$               
 \\   $B_s^{0} \to  Z_c(3900) \eta $   & $5.83^{+1.86}_{-1.86}$      &  $B_s^{0} \to Z_c(4020)\eta $    &~~~~ $0.23^{+0.05}_{-0.05}$    
         \\ 
\hline
\end{tabular}
\end{table*}

One can see that the mechanism we proposed can describe the decays of $B_{(s)}$  into $X(3872)$ plus a strange meson. It is natural to expect that such a mechanism works in similar decays of $B_{(s)}$ into  $Z_c(3900)$ and $Z_{c}(4020)$.  The  decay  constants of $Z_c(3900)$ and $Z_{c}(4020)$ as the isovector $\bar{D}^*D^{(*)}$ molecules are estimated to be $f_{Z_c(3900)}=68.85^{+16.14}_{-21.33}$~MeV and $f_{Z_c(4020)}=15.69^{+2.52}_{-3.01}$~MeV, respectively.   With the decay constants given in Table~\ref{decayconstants12} and the  effective Wilson coefficient $a_2$ given in Table~\ref{resultq2}, we predict the branching fractions of the decays $B \to 
 Z_c(3900) K^*$, $B_s \to Z_c(3900)\eta$, $B_s \to Z_c(3900) \phi$,  $B \to 
 Z_c(4020) K^*$, $B_s \to Z_c(4020)\eta$, and $B_s \to Z_c(4020) \phi$ in Table~\ref{resultsprediction45}, which are smaller than the $B_{(s)}$ decays into $X(3872)$. 

\section{Summary and outlook}
\label{sum}

In summary, we proposed a unified framework to compute the branching fractions of $\bar{D}^*D$ and $\bar{D}^*D^*$ molecules in $B$ decays, where the former molecules refer to  $X(3872)$ and $Z_{c}(3900)$, and the latter to $X_2(4013)$ and $Z_c(4020)$.   Our framework, with no free parameters, 
predicted the branching fractions of  $\mathcal{B}[B^+ \to X(3872)K^+]$ and  $\mathcal{B}[B^0 \to X(3872)K^0]$, $(0.87\sim 2.11)\times 10^{-4}$ and $(0.54 \sim 1.32)\times 10^{-4}$, consistent with the experimental data.  The branching fractions of  $\mathcal{B}[B^+ \to Z_c(3900)K^+]$ and  $\mathcal{B}[B^0 \to Z_c(3900)K^0]$ are found to be about the order of $ 10^{-5}$,  smaller than the experimental upper limits.   Moreover, we predicted the branching fractions of  $\mathcal{B}[B\to X_2(4013) K]$ and $\mathcal{B}[B\to Z_{c}(4020) K]$ to be of the order of $10^{-5}$ and $10^{-6}$.  Simplifying the triangle diagrams as tree-level diagrams, we could extract the decay constants of $XZ$ states as $\bar{D}^*D^{(*)}$ molecules, i.e., $f_{X(3872)}=182.22$ MeV,  $f_{Z_c(3900)}=68.85$ MeV, and $f_{Z_c(4020)}=15.69$ MeV, following the magnitude of the branching fractions of the  $\bar{D}^*D^{(*)}$  molecules in the $B_{(s)}$ decays. With the  $\bar{D}^*D^{(*)}$  molecular decay constants determined, we calculated the branching fractions of the decays $B \to 
 X/Z_c K^*$, $B_s \to  X/Z_c \eta$, and $B_s \to  X/Z_c \phi$. In particular, the calculated branching fractions    $\mathcal{B}[B^+ \to X(3872)K^{*+}]=3.47\times 10^{-4}$  and   $\mathcal{B}[B_s^0 \to X(3872)\phi]=2.51\times 10^{-4}$ are consistent with the current experimental data.

We emphasize that the ratios of branching fractions are more precise than the absolute branching fractions in our framework and can provide more insights into the molecular nature of the states studied. The ratios of     $\mathcal{B}[B^+ \to X(3872) K^+]/\mathcal{B}[B^0 \to X(3872) K^0]$ and $\mathcal{B}[B^+ \to X_2(4013) K^+]/\mathcal{B}[B^0 \to X_2(4013) K^0] $ are about $0.62 $ and $0.75$, the former consistent with the experimental data. The large isospin-breaking effects are attributed to the isospin breaking of the $\bar{D}^*D$ and $\bar{D}^*D^*$ neutral and charged components. On the other hand, the isospin-breaking  ratio  $\mathcal{B}[B^+ \to Z_c(3900) K^+]/\mathcal{B}[B^0 \to Z_c(3900) K^0]=0.63 $ mainly originates from the Wilson coefficient $a_1$ determined by fitting to the weak decay processes of $B^{+(0)}\to D_s^{+} \bar{D}^0( D^{-}) $ and $B^{+(0)}\to D_s^{\ast+} \bar{D}^{\ast0}( D^{\ast-}) $. 
   In addition,  our results show that the branching fractions of $\mathcal{B}[B\to Z_{c}(3900)K] $ are smaller than those of   $\mathcal{B}[B\to X(3872)K]$  by one order of magnitude, which is consistent with the fact  $Z_c(3900)$ has not been observed in  $B$  decays. The predicted hierarchy in the branching fractions of  $\mathcal{B}[B\to Z_{c}(3900)K]$ and $\mathcal{B}[B\to X(3872)K]$  serve as a highly nontrivial test on the molecular nature of $X(3872)$ and $Z_c(3900)$ and should be checked by future experiments. 

A few remarks are in order. In this study, we only considered the dominant  $\bar{D}^*D^{(\ast)}$  contribution to the $XZ$ states. However, other channels, such as $\bar{D}_sD_s$, $\bar{D}^*D^*$, and $\bar{D}_s^*D_s^*$,  can also play a role in forming the $X(3872)$~\cite{Gamermann:2009fv,Ji:2022vdj}. In addition, the $X(3872)$ may contain  a   $c\bar{c}$ component~\cite{Yamaguchi:2019vea}.   The fact that the $\bar{D}^*D$ contribution alone can describe the branching fractions of $X(3872)$ in $B$ decays indicates that $X(3872)$ contains a sizable or dominant $\bar{D}^*D$ component.  As for $Z_c(3900)$, purely based on the $\bar{D}^*D$ invariant mass distributions,  it can also be explained either as a cusp effect or as a virtual state~\cite{Wang:2013cya,Chen:2013coa,Liu:2013vfa,Swanson:2014tra,Albaladejo:2015lob,Pilloni:2016obd,Ortega:2018cnm}, which would affect its couplings to  $\bar{D}^*D$ and therefore modify  $\mathcal{B}[B\to Z_c(3900)K]$. As a result, future experimental measurements of $Z_c(3900)$ in $B$ decays will help either confirm or refute its nature as a $\bar{D}^*D$ resonant state.

\acknowledgments

 M.Z.L  thanks Fu-sheng Yu and Rui-Xiang Shi for useful discussions.    This work is supported in part by the National Natural Science Foundation of China under Grants No.11975041 and No.11961141004. M.-Z.L  acknowledges support from the National Natural Science Foundation of
China under Grant No.12105007.

\appendix

\section{Amplitudes  for weak decays $B\to D_{s}^{(\ast)}\bar{D}^{(\ast)}$ }
\label{appenda}

\begin{figure*}[htbp]
\centering
\includegraphics[width=1.38\columnwidth]{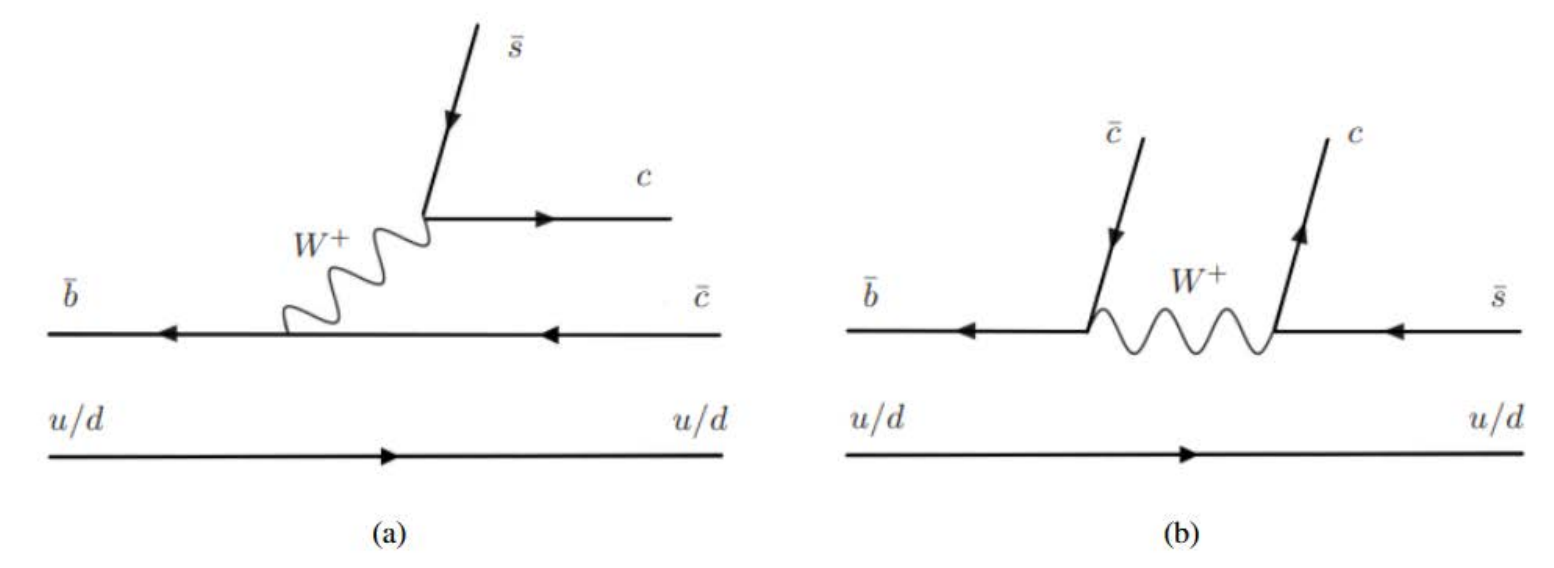}
\caption{External $W$-emission mechanism (a) for $B^{+(0)}\to D_{s}^{(\ast)+} \bar{D}^{(\ast)0}(D^{(\ast)-})$ and internal $W$-emission mechanism (b) for $B^{+(0)} \to J/\psi K^{(*)+(0)}$.    }
\label{quark}
\end{figure*}

The mechanism accounting for the weak decays $B\to D_{s}^{(\ast)} \bar{D}^{(\ast)}$ can be well explained in the naive factorization approach, which   
mainly proceeds via the external $W$-emission mechanism at the quark level shown in Fig.~\ref{quark}~(a).  According to the topological classification of weak decays, the external $W$-emission mechanism often provides the dominant contributions~\cite{Chau:1982da,Chau:1987tk,Molina:2019udw}.    
 As shown in Table~\ref{exp},  the branching fractions $\mathcal{B}[B\to D_{s}^{(\ast)} \bar{D}^{(\ast)}]$ are of the order of $10^{-2}$, and therefore it is favorable to produce the  $\bar{D}^*D^{(*)}$ molecules in $B$ decays via the triangle mechanism.  
 

\begin{table*}[htp]
\centering
\caption{Branching fractions ($10^{-3}$) of  $B\to D_{s}^{(\ast)} \bar{D}^{(\ast)}$. \label{exp}
}
\label{ratios}
\begin{tabular}{|c | c c| c |c c | }
\hline
 &   Decay mode    &~~~~ RPP.~\cite{ParticleDataGroup:2022pth}   &~~ &~~~ Decay mode    &~~~~ RPP.~\cite{ParticleDataGroup:2022pth}  
         \\ \hline
  $a_{1+}^{\prime}$ &~   $B^{+} \to \bar{D}^{0}D_{s}^{+}$       & ~~~~ $9.0\pm 0.9$   & $a_{1+}^{\ast}$ &  ~~           $B^{+} \to \bar{D}^{0}D_{s}^{\ast+} $ & ~~~~ $7.6\pm 1.6$ 
         \\  $a_{10}^{\prime}$ & ~ ${B}^{0} \to {D}^{-}D_{s}^{+}$    & ~~~~ $7.2\pm 0.8$  &$a_{10}^{\ast}$ &~~ ${B}^{0} \to {D}^{-}D_{s}^{\ast+}$    & ~~~~ $7.4\pm 1.6$   
        \\  \hline
          $a_{1+}$ &~~     $B^{+} \to \bar{D}^{\ast0}D_{s}^{+} $ & ~~~~ $8.2\pm 1.7$   &  $a_{1+}^{\prime\ast}$    & ~~~$B^{+} \to \bar{D}^{\ast0}D_{s}^{\ast+} $ & ~~~~ $17.1\pm 2.4$            
\\    $a_{10}$ & ~~${B}^{0} \to {D}^{\ast-}D_{s}^{+}$    & ~~~~ $8.0\pm 1.1$
 & $a_{10}^{\prime\ast}$    & ~~  ${B}^{0} \to {D}^{\ast-}D_{s}^{\ast+}$    & ~~~~ $17.7\pm 1.4$  \\  \hline
\end{tabular}
\end{table*}

 For the weak interaction  vertices, the  decay amplitudes of $B\to  D_{s}^{(\ast)} \bar{D}^{(\ast)}$  can be expressed as the products of two hadronic matrix elements~\cite{Ali:1998eb,Qin:2013tje}
\begin{widetext}
\begin{eqnarray}\label{Ds-KK}
\mathcal{A}\left(B\to D_{s} \bar{D}^{\ast}\right)&=&\frac{G_{F}}{\sqrt{2}} V_{cb}V_{cs} a_{1}\left\langle D_{s}|(s\bar{c})| 0\right\rangle\left\langle \bar{D}^{\ast }|(c \bar{b})| B\right\rangle , \\
\mathcal{A}\left(B\to D_{s}\bar{D}\right)&=&\frac{G_{F}}{\sqrt{2}} V_{cb}V_{cs} a_{1}^{\prime}\left\langle D_{s}|(s\bar{c})| 0\right\rangle\left\langle \bar{D}^{ }|(c \bar{b})| B\right\rangle , \\
\mathcal{A}\left(B\to D_{s}^{\ast} \bar{D}\right)&=&\frac{G_{F}}{\sqrt{2}} V_{cb}V_{cs} a_{1}^*\left\langle D_{s}^{\ast}|(s\bar{c})| 0\right\rangle\left\langle \bar{D}|(c \bar{b})| B\right\rangle , \\ 
\mathcal{A}\left(B\to D_{s}^{\ast} \bar{D}^{\ast}\right)&=&\frac{G_{F}}{\sqrt{2}} V_{cb}V_{cs} a_{1}^{\prime\ast}\left\langle D_{s}^{\ast}|(s\bar{c})| 0\right\rangle\left\langle \bar{D}^{\ast }|(c \bar{b})| B\right\rangle ,
\end{eqnarray}
\end{widetext}
where $a_{1}=c_{1}^{e f f}+c_{2}^{e f f} / N_{c}$ with $N_c$ the number of colors and $a_1$  can be obtained in the factorization approach~\cite{Bauer:1986bm}.  In the present work, we determine $a_1$ by fitting to the experimental branching fractions.

 The current matrix elements between a pseudoscalar meson or vector meson and the vacuum have the following form:

\begin{eqnarray}
\left\langle D_{s}|(s \bar{c})| 0\right\rangle  =i f_{D_{s}} p^{\mu}_{D_{s}},  \\ \nonumber
\left\langle D_{s}^{\ast} |(s\bar{c} )| 0\right\rangle = m_{D_{s}^{\ast}}f_{D_{s}^{\ast}}\epsilon_\mu^*, 
\end{eqnarray}
where $f_{D_{s}}$ and  $f_{D_{s}^{\ast}}$  are the decay constants for $D_s$ and $D_s^{\ast}$, respectively, and $\epsilon_\mu^*$ denotes the polarization vector of a vector particle.  In this work, we take  $G_F = 1.166 \times 10^{-5}~{\rm GeV}^{-2}$, $V_{cb}=0.041$, $V_{cs}=0.987$, $f_{D_{s}} = 250$ MeV, and $f_{D_{s}^{\ast}}=272$~MeV~\cite{ParticleDataGroup:2022pth,Verma:2011yw,FlavourLatticeAveragingGroup:2019iem,Li:2017mlw}.  
 
The hadronic matrix elements are parameterized in terms of six form factors~\cite{Verma:2011yw}
\begin{widetext}
\begin{eqnarray}
&&\left\langle \bar{D}^{\ast}|(c\bar{b})| B\right\rangle=\epsilon_{\alpha}^{*}\left\{-g^{\mu \alpha} (m_{\bar{D}^{\ast}}+m_{B}) A_{1}\left(q^{2}\right)+P^{\mu} P^{\alpha} \frac{A_{2}\left(q^{2}\right)}{m_{\bar{D}^{\ast}}+m_{B}}\right. \\  \nonumber
&&+i \varepsilon^{\mu \alpha \beta \gamma}P_\beta q_\gamma \left.\frac{V\left(q^{2}\right)}{m_{\bar{D}^{\ast}}+m_{B}} +q^{\mu} P^{\alpha} \left[\frac{m_{\bar{D}^{\ast}}+m_{B}}{q^{2}}A_{1}\left(q^{2}\right)-\frac{m_{B}-m_{\bar{D}^{\ast}}}{q^{2}}A_{2}\left(q^{2}\right)-\frac{2m_{\bar{D}^{\ast}}}{q^{2}}A_{0}\left(q^{2}\right)\right]\right\},
\\ 
&&\left\langle \bar{D}|(c \bar{b} )| B\right\rangle =\left[(p_{B}+p_{\bar{D}})^{\mu}-\frac{m_{B}^2-m_{\bar{D}}^2}{q'^2}q'_{\mu}\right] F_{1}(q'^2)+\frac{m_{B}^2-m_{\bar{D}}^2}{q'^2}q'_{\mu} F_{0}(q'^2),
\end{eqnarray}
\end{widetext}
where $q$ and  $q^{\prime}$   represent the momentum transfer of $ p_{B}- p_{\bar{D}^{\ast}}$ and $ p_{B}- p_{\bar{D}}$,  respectively, and $P = p_{B} + p_{\bar{D}^{\ast}}$.  

The form factors of  $F_{1,0}(t)$,  $A_{0}(t)$, $A_{1}(t)$, $A_{2}(t)$, and $V(t)$ with $t \equiv q^{(\prime)2}$ can be parameterized in the following form~\cite{Verma:2011yw} 
\begin{equation}
F(t)=\frac{F(0)}{1-a\left(t / m_{B}^{2}\right)+b\left(t^{2} / m_{ B}^{4}\right)}.
\end{equation}
The values of $F_0$, $a$, and $b$ in  the transition form factors of $B\to \bar{D}^{(\ast)}$ are taken from Ref.~\cite{Verma:2011yw} and shown in Table~\ref{BtoKformfactor1}. 
  
  The  weak decay amplitudes of { $B(k_0) \to D_{s}^{(\ast)}(q_1)\bar{D}^{(\ast)}(q_2)$ }have the following form
 \begin{widetext} 
\begin{align}\label{am3}
\mathcal{A}(B\to D_{s}\bar{D}^{\ast})&= \frac{G_{F}}{\sqrt{2}}V_{cb}V_{cs} a_{1} f_{D_{s}}\{-q_{1}\cdot \varepsilon(q_{2})(m_{\bar{D}^{\ast}}+m_{B}) A_{1}\left(q_{1}^{2}\right)   \\ \nonumber 
&+(k_{0}+q_{2}) \cdot \varepsilon(q_{2}) q_{1}\cdot (k_{0}+q_{2}) \frac{A_{2}\left(q_{1}^{2}\right)}{m_{\bar{D}^{\ast}}+m_{B}} +(k_{0}+q_{2}) \cdot \varepsilon(q_{2}) \\ \nonumber       & 
[(m_{\bar{D}^{\ast}}+m_{B})A_{1}(q_{1}^2) -(m_{B}-m_{\bar{D}^{\ast}})A_{2}(q_1^2) -2m_{\bar{D}^{\ast}} A_{0}(q_{1}^2)]  \} , \\ \nonumber
\mathcal{A}(B\to D_{s}\bar{D})&=\frac{G_{F}}{\sqrt{2}}V_{cb}V_{cs} a_{1}^{\prime}f_{D_{s}}(m_{B}^2-m_D^2)F_{0}(q_{1}^2),
\\ \nonumber
\mathcal{A}(B\to D_{s}^{\ast}\bar{D})&= \frac{G_{F}}{\sqrt{2}}V_{cb}V_{cs} a_{1}^{*}m_{D_{s}^{\ast}}f_{D_{s}^{\ast}}(k_{0}+q_{2})\cdot \varepsilon(q_{1})F_{1}(q_{1}^2), \\ \nonumber
 \mathcal{A}(B\to D_{s}^{\ast}\bar{D}^{\ast})&= \frac{G_{F}}{\sqrt{2}}V_{cb}V_{cs} a_{1}^{*\prime}m_{D_{s}^{\ast}}f_{D_{s}^{\ast}}\varepsilon_{\mu}(q_1)\left[(-g^{\mu \alpha} (m_{\bar{D}^{\ast}}+m_{B}) A_{1}\left(q_1^{2}\right)\right.  \\ \nonumber &+ \left. (k_0+q_2)^{\mu} (k_0+q_2)^{\alpha} \frac{A_{2}\left(q_1^{2}\right)}{m_{\bar{D}^{\ast}}+m_{B}}  
+i \varepsilon^{\mu \alpha \beta \gamma}(k_0+q_2)_\beta q_{1\gamma} \frac{V\left(q_1^{2}\right)}{m_{\bar{D}^{\ast}}+m_{B}}\right]\varepsilon_{\alpha}(q_{2}).
\end{align}
\end{widetext}
By fitting to the eight branching fractions of $B\to D_{s}^{(\ast)}\bar{D}^{(\ast)}$ tabulated in Table~\ref{ratios},  we obtain  $a_{1+}=0.929$, $a_{10}=0.955$,      $a_{1+}^{\prime}=0.791$, $a_{10}^{\prime}=0.736$,  $a_{1+}^{ \ast}=0.812$, $a_{10}^{\ast}=0.834$,  $a_{1+}^{\prime\ast}=0.833$, and $a_{10}^{\prime\ast}=0.880$. The subscript with $+$ and $0$ denote the $B^+$ and $B^0$ decay modes, the values of which show small isospin-breaking effects of less than 10\%.

{     The  weak decay amplitudes of $B (k_0)  \to X(3872) (p_2) K^{(*)}(p_1)$ have the following form
 \begin{widetext} 
\begin{align}\label{am3}
\mathcal{A}(B\to X(3872)K )&= \frac{G_{F}}{\sqrt{2}}V_{cb}V_{cs} a_{2}m_{X(3872)}f_{X(3872)}(k_{0}+p_1)\cdot \varepsilon(p_2)F_{1}(p_2^2),  \\ \nonumber
\mathcal{A}(B\to X(3872)K^{\ast})&=  \frac{G_{F}}{\sqrt{2}}V_{cb}V_{cs} a_{2} m_{X(3872)}f_{X(3872)}\varepsilon_{\mu}(p_2)\left[(-g^{\mu \alpha} (m_{X(3872)}+m_{B}) A_{1}\left(p_2^{2}\right)\right.  \\ \nonumber &+ \left. (k_0+p_1)^{\mu} (k_0+p_1)^{\alpha} \frac{A_{2}\left(p_2^{2}\right)}{m_{X(3872)}+m_{B}}  
+i \varepsilon^{\mu \alpha \beta \gamma}(k_0+p_1)_\beta p_{2\gamma} \frac{V\left(p_2^{2}\right)}{m_{X(3872)}+m_{B}}\right]\varepsilon_{\alpha}(p_1).
\end{align}
\end{widetext}

}

 \begin{table}[ttt]
 \centering
 \caption{Values of  $F(0)$, $a$, $b$ in the $B \rightarrow K$ and $B  \rightarrow K^*$  transition  form factors~\cite{Verma:2011yw}. \label{BtoKformfactor1} }
 \begin{tabular}{|c|cccccc|}
 \hline
  & $F_0$~~~ & $F_1$~~~ & $V$~~~ & $A_0$~~~ &  $A_1$~~~ & $A_2$~~~\\
 \hline  
 $F(0)^{B \to D^{(*)}}$~~~ &  0.67~~~  & 0.67~~~ & 0.77~~~ & 0.68~~~ & 0.65~~~ & 0.61~~~\\
 $a^{B \to D^{(*)}}$~~~  & 0.63~~~ &  1.22~~~  & 1.25~~~ & 1.21~~~ & 0.60~~~ & 1.12~~~\\
 $b^{B \to D^{(*)}}$~~~ & 0.01~~~  & 0.36~~~  & 0.38~~~ &0.36~~~ & 0.00~~~ & 0.31~~~\\ \hline
 $F(0)^{B\to K^{(*)}}$~~~ &  0.34~~~  & 0.34~~~ & 0.36~~~ & 0.38~~~ & 0.31~~~ & 0.28~~~\\
 $a^{B\to K^{(*)}}$~~~  & 0.78~~~ &  1.60~~~  & 1.69~~~ & 1.61~~~ & 0.84~~~ & 1.53~~~\\
 $b^{B\to K^{(*)}}$~~~ & 0.05~~~  & 0.73~~~  & 0.95~~~ &0.89~~~ & 0.12~~~ & 0.79~~~\\
  \hline
 $F(0)^{B_{s}\to \eta(\phi)}$~~~ &  0.28~~~  & 0.28~~~ & 0.29~~~ & 0.31~~~ & 0.25~~~ & 0.22~~~\\
 $a^{B_{s}\to \eta(\phi)}$~~~  & 1.07~~~ &  1.82~~~  & 1.95~~~ & 1.87~~~ & 1.20~~~ & 1.79~~~\\
 $b^{B_{s}\to \eta(\phi)}$~~~ & 0.32~~~  & 1.45~~~  & 1.98~~~ &1.87~~~ & 0.54~~~ & 1.67~~~\\
\hline
 \end{tabular}
 \end{table}

\section{Amplitudes for $D_{s}^{(*)}\to D^{(*)}K$   }
\label{appendb}

  The Lagrangians  describing the interactions between charmed mesons and the kaon 

\begin{eqnarray}
\mathcal{L}_{D_{s}^{\ast} D K}&=& -i g_{D_{s}^{\ast} D K} ( D \partial^{\mu} K D_{s\mu}^{\ast\dag}-  D_{s\mu}^{\ast}\partial^{\mu} K D^{\dag}), \\ \nonumber
\mathcal{L}_{D_{s} D^{\ast} K}&=& -i g_{D_{s} D^{\ast} K} (D_{s} \partial^{\mu} K D^{\ast\dag}_{\mu}-D_{\mu}^* \partial^{\mu} K  D_{s}^{\dag}), \\ \nonumber
\mathcal{L}_{D_{s}^{\ast} D^{\ast} K}&=& - g_{D_{s}^{\ast} D^{\ast} K} \varepsilon_{\mu\nu\alpha\beta} \partial^{\mu}D_{s}^{\ast\nu} {\partial}^{\alpha} {D}^{\ast\beta\dag} K , 
\end{eqnarray}
where $g_{D_{s} D^{\ast} K}$, $g_{D_{s}^{\ast} D K}$, and $ g_{D_{s}^{\ast} D^{\ast} K}$ are the couplings to be determined.  From these Lagrangians, one can obtain the amplitudes for the decays of {  ${D}_{s}^{(\ast)}(q_1)\to {D}^{(\ast)}(q_3)K(p_1)$ }

\begin{eqnarray}
\mathcal{A}\left({D}_{s}^{\ast} \to {D} K \right)&=&g_{{D}_{s}^{\ast} D K}p_{1}\cdot \varepsilon(q_{1}),\\     \nonumber
\mathcal{A}\left(D_{s} \to {D}^{\ast} K \right)&=&-g_{D_{s} {D}^{\ast} K}p_{1}\cdot \varepsilon(q_{3}),  \\  \nonumber
  \mathcal{A}\left({D}_{s}^*\to {D}^{\ast} K \right) &=& g_{{D}_{s}^*  {D}^{\ast} K}\varepsilon_{\mu\nu\alpha\beta}q_{1}^{\mu}\varepsilon^{\nu}(q_{1})q_{3}^{\alpha}\varepsilon^{\beta}(q_{3}). 
\end{eqnarray}
Assuming SU(3)-flavor symmetry and SU(4)-flavor symmetry the coupling $g_{D_{s} D^{\ast} K}$ is estimated to be  $16.6$~\cite{Xie:2022lyw} and $10$~\cite{Azevedo:2003qh}, while the QCD sum rule yields $5$~\cite{Bracco:2006xf,Wang:2006ida}. 
 In view of this large variance, we adopt the couplings estimated by SU(4) symmetry, which are in between those estimated utilyzing SU(3) symmetry and by the QCD sum rule, i.e.,  $g_{D_{s} D^{\ast} K}=g_{D_{s}^{\ast} D K}=10$ and  $g_{D_{s}^{\ast} D^{\ast} K}=7.0$~GeV$^{-1}$~\cite{Azevedo:2003qh}. It is well known that the SU(4) symmetry is heavily broken. Therefore, following Refs.~\cite{Follana:2007uv,Carrasco:2014poa}, we estimate the breaking of SU(4) symmetry using the $D$ and $K$ decay constants as references, which is $33\%$, consistent with Ref.~\cite{Fontoura:2017ujf}.

\section{Amplitudes for  the dynamical generation of $\bar{D}^*D$ and $\bar{D}^*D^*$ molecules}
\label{appendc}

The Lagrangian describing the interactions between $X(3872)$/$X_{2}(4013)$ and their constituents are 

\begin{eqnarray}
\mathcal{L}_{X \bar{D} D^{\ast}}&=& g_{X \bar{D} D^{\ast}} X^{\mu}  {D}^{\ast}_{\mu} \bar{D},    \\ \nonumber
\mathcal{L}_{X_2 \bar{D}^* D^{\ast}}&=& g_{X_2 \bar{D}^* D^{\ast}} X_{2\mu\nu}{D}^{\ast\mu} \bar{D}^{\ast\nu},  
\end{eqnarray}
where $g_{X \bar{D} D^{\ast}}$ and $ g_{X_2 \bar{D}^* D^{\ast}}$ are  the couplings to be determined. 

In the following, we denote $Z_{c}(3900)$ and $Z_{c}(4020)$ as $Z_c$ and $Z_c^{\prime}$, respectively. 
The interactions between the $Z_{c}(3900)/Z_{c}(4020)$ resonant states  and their constituents can be expressed by the following effective Lagrangians:
\begin{eqnarray}
\mathcal{L}_{Z_{c}D\bar{D}^*}&=&  g_{Z_{c}D\bar{D}^*} Z_{c}^{\mu}D\bar{D}^*_{\mu},  \\ \nonumber
\mathcal{L}_{Z_{c}^{\prime}D^*\bar{D}^*}&=& i g_{Z_{c}^{\prime}D^*\bar{D}^*}\varepsilon_{\mu\nu\alpha\beta} \partial^{\mu}Z_{c}^{\prime\nu}D^{\ast\alpha}\bar{D}^{\ast\beta},
\end{eqnarray}
where $g_{Z_{c}D\bar{D}^*} $ and $g_{Z_{c}^{\prime}D^*\bar{D}^*}$ are the corresponding couplings.

From the above Lagrangians, one can obtain the amplitudes for the coupling of $X(3872)$, $X_2(4013)$, $Z_c(3900)$, $Z_c(4020)$ to the $\bar{D}D^*$ and  $\bar{D}D^*$  channels, e.g., {  $ \bar{D}^{(*)} (q_2)D^{(*)} (q_3) \to X/Z_c (p_2)$ }
\begin{widetext}
\begin{eqnarray}
\mathcal{A}(\bar{D}D^*\to X(3872)) &=& g_{\bar{D}D^*  X} \varepsilon^{\mu}(p_2)\varepsilon_{\mu}(q_3), \\
\mathcal{A}(\bar{D}^*D\to X(3872)) &=& g_{\bar{D}D^*  X} \varepsilon^{\mu}(p_2)\varepsilon_{\mu}(q_2), \\
\mathcal{A}(\bar{D}^*D^*\to X_2(4013)) &=& g_{\bar{D}^*D^*  X} \varepsilon^{\mu\nu}(p_{2})\varepsilon_{\mu}(q_{3})\varepsilon_{\nu}(q_{2}),  \\
\mathcal{A}(\bar{D}D^*\to Z_{c}(3900)) &=& g_{\bar{D}D^* Z_{c}} \varepsilon^{\mu}(p_2)\varepsilon_{\mu}(q_3), \\
\mathcal{A}(\bar{D}^*D\to  Z_{c}(3900)) &=& g_{\bar{D}D^*  Z_{c}} \varepsilon^{\mu}(p_2)\varepsilon_{\mu}(q_2), \\
\mathcal{A}(\bar{D}^*D^*\to Z_{c}(4020)) &=& g_{\bar{D}^*D^*  Z_{c}^{\prime}} \varepsilon_{\mu\nu\alpha\beta}p_{2}^{\mu}\varepsilon^{\nu}(p_{2})\varepsilon^{\alpha}(q_3)\varepsilon^{\beta}(q_{2}). 
\end{eqnarray}
\end{widetext}
{
In the above formula, 
$\varepsilon^{\mu}(q)$ denotes the polarization vector of a particle with spin $S=1$, and  $\varepsilon^{\mu\nu}(q)$ denotes the  polarization tensor of a particle with spin $S=2$.}

The above couplings are determined by the contact-range effective field theory as shown later. One should note that the isovector $\bar{D}^*D^{(*)}$ molecules are generated by the $\bar{D}^*D^{(*)}$ potentials in isospin basis, and therefore their couplings to the components in particle basis can be derived following Eq.~(\ref{eqisovector}).  
 In the isospin limit, one can decompose the isospin wave functions of the isovector molecules as  
 \begin{widetext}
\begin{eqnarray}
\label{eqisovector}
| Z_{c}(3900) \rangle &=& \frac{1}{2} \left[(| {D}^{\ast+}D^- \rangle -| {D}^{\ast0}\bar{D}^{0} \rangle)+ (|{D}^{+}D^{\ast-} \rangle -| D^{0} \bar{D}^{\ast0}\rangle)\right],  \\ \nonumber
| Z_{c}(4020) \rangle &=& \frac{1}{\sqrt{2}} \left(| {D}^{\ast+}D^{\ast-} \rangle -| {D}^{\ast0}\bar{D}^{\ast0} \rangle\right). 
\end{eqnarray}
 \end{widetext}
  To compare with the isovector molecules, 
we  decompose the isospin wave functions of  the isoscalar molecules as 
\begin{widetext}
\begin{eqnarray}
\label{eqisocalar}
| X(3872) \rangle &=& \frac{1}{2} \left[(| {D}^{\ast+}D^- \rangle +| {D}^{\ast0}\bar{D}^{0} \rangle)- (|{D}^{+}D^{\ast-} \rangle +| D^{0}\bar{D}^{\ast0} \rangle)\right],  \\ \nonumber
| X_2(4013) \rangle &=& \frac{1}{\sqrt{2}} \left(| {D}^{\ast+}D^{\ast-} \rangle +| {D}^{\ast0}\bar{D}^{\ast0} \rangle\right).
\end{eqnarray}
 \end{widetext}
Due to the fact that the isoscalar $\bar{D}^*D^{(*)}$ molecules  are generated by the $\bar{D}^*D^{(*)}$ potentials in particle basis,  the absolute couplings of the isoscalar $\bar{D}^*D^{(*)}$  molecules   to their components   are determined by the EFT approach, 
but the sign between the components is determined by Eq.~(\ref{eqisocalar}).

\section{Contact-range effective field theory for the $\bar{D}^*D$ and $\bar{D}^*D^*$ interactions}
\label{appendd}

In this subsection, we briefly describe the contact-range effective field theory in which the $\bar{D}^*D$ and $\bar{D}^*D^*$ interactions can dynamically generate the $X(3872)$, $X_2(4013)$, $Z_c(3900)$, and $Z_c(4020)$. 

The scattering amplitude $T$ responsible for the dynamical generation of the 
  $\bar{D}^*D$ and $\bar{D}^*D^*$ molecules  is determined by   solving the Lippmann-Schwinger equation 
\begin{eqnarray}
T=(1-VG)^{-1}V,
\end{eqnarray}
where $V$ is the $\bar{D}^{(*)}D^{(*)}$ potential determined by the contact-range EFT approach shown later, and $G$ is the two-body propagator. 
In evaluating the loop function $G$, we introduce a regulator of Gaussian form $e^{-2q^{2}/\Lambda^2}$ in the integral as
\begin{eqnarray}
G(s)=\int \frac{d^{3}q}{(2\pi)^{3}} \frac{e^{-2q^{2}/\Lambda^2}}{{\sqrt{s}}-m_{1}-m_{2}-q^{2}/(2\mu_{12})+i \varepsilon}
\label{loopfunction},
\end{eqnarray}
where $\sqrt{s}$ is  the total energy in the c.m. frame of $m_{1}$ and $m_{2}$, $\mu_{12}=\frac{m_{1}m_{2}}{m_{1}+m_{2}}$ is the reduced mass, and $\Lambda$ is the momentum cutoff. Following our previous works~\cite{Liu:2019tjn,Xie:2022hhv}, we take $\Lambda=1$~GeV in the present work.        The dynamically generated states correspond to poles in the unphysical sheet.  In this sheet, the loop function of Eq.~(\ref{loopfunction}) becomes
\begin{eqnarray}
G^{II}(s,m_1,m_2)=G^{I}(s,m_1,m_2)+i\mu_{12} \frac{p  }{2\pi}{  e^{-2 p^2/\Lambda^2}},
\label{loopfunction1}
\end{eqnarray}
where the c.m. momentum $p$  is  
\begin{eqnarray}
p=\sqrt{2\mu_{12}\left(\sqrt{s}-m_1-m_2\right)}.
\end{eqnarray}

Using   the $\bar{D}^{(*)}D^{(*)}$  contact potentials  we can search for  poles  around  the $\bar{D}^{(*)}D^{(*)}$   thresholds, and  determine the 
 couplings between the molecular states and their constituents from the residues of the corresponding poles, 
\begin{eqnarray}
g_{i}g_{j}=\lim_{{\sqrt{s}}\to {\sqrt{s_0}}}\left({\sqrt{s}}-{\sqrt{s_0}}\right)T_{ij}(\sqrt{s}),
\end{eqnarray}
where $g_{i}$ denotes the coupling of channel $i$ to the  dynamically generated state and ${\sqrt{s_0}}$ is the pole position.

In the heavy quark limit, the contact potential of the $J^{PC}=1^{++}$ $\bar{D}^*D$ channel is expressed as a sum of two low-energy constants, $C_a+C_b$, where $C_a$ characterizes the spin-independent interaction and $C_b$ accounts for the spin-spin interaction~\cite{Nieves:2012tt}. Because the $X(3872)$ is quite close to the mass threshold of $\bar{D}^{\ast0}D^{0}$ but below the mass threshold of $D^{\ast+}D^{-}$ by $8$~MeV, it is important to take into account the isospin-breaking effects of the $\bar{D}^*D$ wave function. It contains both a neutral component and a charged component, which are defined as $C_{n}=\frac{1}{\sqrt{2}}(\bar{D}^{\ast0}D^0-\bar{D}^{0}D^{\ast0})$  and $C_{c}=\frac{1}{\sqrt{2}}(D^{\ast+}D^--{D}^{+}D^{\ast-})$. The contact potential for $C_n$ and $C_c$ read~\cite{Nieves:2012tt}:  
\begin{equation}
    V_{C_{n}-C_{c}}=\begin{pmatrix}C_a^{\prime}+C_{b}^{\prime}& C_a^{\prime\prime}+C_{b}^{\prime\prime}
    \\ C_a^{\prime\prime}+C_{b}^{\prime\prime} & C_a^{\prime}+C_{b}^{\prime}\end{pmatrix}.
\end{equation}

In principle, the contact potentials should be determined by reproducing the experimental data. Due to the scarcity of experimental data one needs to turn to other approaches such as the light meson saturation mechanism, which dictates that the couplings are saturated by the light meson ($\sigma$, $\rho$, and $\omega$) exchanges in the one boson exchange model. $C_a$ receives contributions from both the scalar and vector meson exchanges, but $C_b$ only receives contributions from the vector meson exchanges, i.e.,
\begin{eqnarray}
C_{a}^{S}+ C_{a}^{V}&\propto& C_{a}^{sat}(\Lambda\sim m_{\sigma},m_{V}),    \\ \nonumber
C_{b}^{V}&\propto & C_{b}^{sat}(\Lambda\sim m_{\sigma},m_{V}),  
\end{eqnarray}
where  
\begin{eqnarray}
C_{a}^{sat(\sigma)}(\Lambda\sim m_{\sigma})&\propto& -\frac{g_{\sigma}^2}{m_{\sigma}^2},    \\ \nonumber
C_{a}^{sat (V)}(\Lambda\sim m_{V})&\propto& -\frac{g_{v}^2}{m_{v}^2}(1-\vec{\tau}_{1}\cdot\vec{\tau}_{2}),  \\ \nonumber
C_{b}^{sat (V)}(\Lambda\sim m_{V})&\propto& -\frac{f_{v}^2}{6 M^2}(1-\vec{\tau}_{1}\cdot\vec{\tau}_{2}),
\label{123}
\end{eqnarray}
where $g_{\sigma_{1}}$ denotes the coupling of the charm meson to the sigma meson, $g_{v1}$ and $f_{v1}$ denote the electric-type and magnetic-type couplings between the charm meson and a light vector meson, and $M$ is a mass scale to render $f_{v1}$ dimensionless.
The proportionality constant is unknown and depends on the details of the renormalization procedure. However, assuming that the constant is the same for $C_{a}^{sat}$ and $C_{b}^{sat}$, we can calculate their ratio. Such an approach has been verified in the studies of the $\bar{D}^{(*)}\Sigma_c^{(*)}$ system and it was found that the ratio estimated by the light meson saturation is consistent with that obtained by reproducing the masses of the $P_{c}$ states~\cite{Liu:2019zvb}. 

As for the $C_n-C_c$ system,  the diagonal potential receives contributions from the $\sigma$, $\omega$, and $\rho^0$ mesons but the off-diagonal potentials only receive contributions from the $\rho^{+}$ meson, which can determine the relative strength between the off-diagonal potential and the diagonal potential.  Following Ref.~\cite{Peng:2020xrf}, the couplings of $C_a^{\prime}$, $C_b^{\prime}$, $C_a^{\prime\prime}$, and $C_b^{\prime\prime}$ in the light meson saturation approach read
\begin{eqnarray}
C_{a}^{\prime}&\propto& -\frac{g_{\sigma}^2}{m_{\sigma}^2} -\frac{g_{v}^2}{m_{v}^2}-\frac{g_{v}^2}{m_{v}^2}, \\ \nonumber
C_{b}^{\prime}&\propto&  -\frac{f_{v}^2}{6M^2}-\frac{f_{v}^2}{6M^2},  
\\ \nonumber
C_{a}^{\prime\prime}&\propto& -2\frac{g_{v}^2}{m_{v}^2}, \\ \nonumber
C_{b}^{\prime\prime}&\propto&  -2\frac{f_{v}^2}{6M^2},
\end{eqnarray}
from which we obtain the ratio of  $C_a^{\prime\prime}+C_b^{\prime\prime}$ to $C_a^{\prime}+C_b^{\prime}$:
\begin{eqnarray}
\frac{C_a^{\prime\prime}+C_b^{\prime\prime}}{C_a^{\prime}+C_b^{\prime}} \approx 0.5,
\end{eqnarray}
consistent with the estimations of Refs.~\cite{Hidalgo-Duque:2012rqv,Albaladejo:2015dsa,Ji:2022uie}. As a result, the unknown couplings of the contact-range potentials are reduced to one. By reproducing the mass of $X(3872)$,  we obtain  $C_a^{\prime}+C_b^{\prime}=-12.551$~GeV$^{-2}$ and the corresponding couplings to the neutral and charged components of its wave function, i.e., $g_{n}=3.86$~GeV and $g_{c}=3.39$~GeV.   Taking into account HQSS,  one can obtain the potentials of the $\bar{D}^{*0}D^{\ast0}/D^{*+}D^{\ast-}$ system and predict the existence of a $J^{PC}=2^{++}$ bound state with a mass of $m=4013.03$~MeV, corresponding to $X_2(4013)$. Finally, the $X_2(4013)$ couplings to its neutral and charged components are  determined as $g_{n}^{\prime}=5.36$~GeV and    $g_{c}^{\prime}=4.86$~GeV.

\begin{table}[ttt]
\centering
\caption{ Values of the couplings of the $\bar{D}^*D^{(*)}$ molecules to their neutral and charged components. \label{couplingparticle}
}
\begin{tabular}{|c| c c |c c c c c c}
\hline
 Molecules     &~~$g_n$    &~~~~~~~$g_c$
         \\ \hline
        $X(3872) $   &~~ $3.86$~GeV   &~~~~~~~~$3.39$~GeV    \\
        $X_2(4013)$   &~~~$5.36$~GeV     &~~~~~~~~$4.86$~GeV  
         \\  
        $Z_c(3900) $   &~~~$5.02$~GeV  &~~~~~~~~$5.02$~GeV   \\
        $Z_c(4020)$  &~~$1.25$  &~~~~~~~~$1.25$ 
\\
\hline
\end{tabular}
\end{table}

 To generate resonant states, the contact potential has to be supplemented with a $q^2$ dependent term~\cite{Yang:2020nrt}, where $q$ is the relative three momentum. Identifying $Z_c(3900)$ as a $\bar{D}D^{\ast}$ resonant state  with the form  of $C_{s}+C_{d}~q^2$,  we  obtain  $C_{s}=-7.7$ GeV$^{-2}$ and  $C_{d}=-211$ GeV$^{-4}$, and then   the coupling 
$g_{Z_{c}(3900)\bar{D}D^*}=7.10$~GeV. Taking into account HQSS,  we predict the existence of a $\bar{D}^*D^*$ molecule with  $M=4028$ MeV and $\Gamma=26$~MeV,  in perfect agreement with the experimental measurements, and then   
obtain the coupling $g_{Z_{c}(4020)\bar{D}^*D^*}=1.77$.  In Table~\ref{couplingparticle}, we collect the values of the molecular couplings.

\section{Additional numerical details }
\label{appende}

\begin{table*}[htp]
\centering
\caption{Branching fractions ($10^{-4}$) of $B^{+(0)}\to X(3872)/X_2(4013) K^{+(0)}$ and $B^{+(0)}\to Z_{c}(3900)/Z_{c}(4020) K^{+(0)}$ and ratios $\mathcal{B}(B^0\to)$/$\mathcal{B}(B^+\to)$. \label{resultsd}
}
\begin{tabular}{|c| c c| c c | }
\hline
    Decay modes    &~~~~ Our predictions  &~~~~ Exp.~\cite{ParticleDataGroup:2022pth}   &~~~ Ratio ~~~~   &    ~~~~ Exp. data~\cite{ParticleDataGroup:2022pth}
         \\ \hline 
        $B^{+} \to X(3872)K^+ $   &~~~~ $1.49\pm 0.62$   &~~~~ $2.1\pm 0.7$   & ~~~\multirow{2}{2.2cm}{ $0.62\pm0.13$}   &~~~~~\multirow{2}{1.8cm}{ $0.52\pm 0.26$}
                 \\   $B^{0} \to X(3872)K^0 $   &~~~~ $0.93\pm0.39$  &~~~~ $1.1 \pm 0.4$  
            & &
         \\ \hline  $B^{+} \to X_2(4013)K^+ $  &~~~~ $0.23\pm0.08$    &~~~~$-$   & ~~~\multirow{2}{2.2cm}{ $0.75\pm0.16$}  &~~~~~\multirow{2}{1.0cm}{ $-$}
         \\     $B^{0} \to X_2(4013)K^0 $  &~~~~ $0.17\pm0.06$   &~~~~ $-$  
         &   &
             \\\hline
        $B^{+} \to Z_{c}(3900)K^+ $   &~~~~ $0.21\pm 0.11$   &~~~~ $<4.7\times 10^{-5}$   & ~~~\multirow{2}{2.2cm}{ $0.63 \pm0.29$} &~~~~~\multirow{2}{1.0cm}{ $-$}
                 \\   $B^{0} \to Z_{c}(3900)K^0 $   &~~~~ $0.13\pm0.07$  &~~~~ $-$  & & 
      \\ \hline
   $B^{+} \to Z_{c}(4020)K^+ $  &~~~~  $0.0095\pm 0.0033$  &~~~~$<1.6\times 10^{-5} $   & ~~~\multirow{2}{2.2cm}{ $1.05\pm0.14$} &~~~~~\multirow{2}{1.0cm}{ $-$}
         \\     $B^{0} \to Z_{c}(4020) K^0 $  &~~~~ $0.0100\pm 0.0034$    &~~~~ $-$   & &   
         \\  
\hline
\end{tabular}
\end{table*}

In Table~\ref{resultsd}, we present the branching fractions of the decays  of $B\to X(3872)/X_2(4013) K$ and $B\to Z_{c}(3900)/Z_{c}(4020) K$.  In the following, we analyze the origin of the isospin breaking in the ratios $\mathcal{B}[B^0\to X(3872)K^0]/\mathcal{B}[B^+\to X(3872)K^+]$ and  $\mathcal{B}[B^+\to Z_{c}(3900)K^+]/\mathcal{B}[B^0\to Z_{c}(3900)K^0]$.

In the particle basis,  the Wilson coefficients $a_1^{\prime}/a_{1}^{\prime *}$ and the couplings $g_n/g_c$  for the decays of $B^0\to X(3872)K^0$ and $B^+\to X(3872)K^+$ are different, resulting in a ratio  $\mathcal{B}[B^0\to X(3872)K^0]/\mathcal{B}[B^+\to X(3872)K^+]=0.62\pm 0.13$. With different couplings $g_n/g_c$  but the same  Wilson coefficients $a_1^{\prime}/a_{1}^{\prime *}$, the ratio becomes  $\mathcal{B}[B^0\to X(3872)K^0]/\mathcal{B}[B^+\to X(3872)K^+]=0.66\pm 0.14$.  On the other hand, with the same couplings $g_n/g_c$ but different Wilson coefficients $a_1^{\prime}/a_{1}^{\prime *}$, the ratio  becomes $\mathcal{B}[B^0\to X(3872)K^0]/\mathcal{B}[B^+\to X(3872)K^+]=0.81\pm 0.17$.  Clearly,  the isospin-breaking of the  ratio  $\mathcal{B}[B^0\to X(3872)K^0]/\mathcal{B}[B^+\to X(3872)K^+]$ is mainly caused by the isospin breaking of the $\bar{D}^*D$ wave function.

For the $Z_c(3900)$, the different  Wilson coefficients $a_1^{\prime}/a_{1}^{\prime *}$ lead to   the ratio  $\mathcal{B}[B^+\to Z_{c}(3900)K^+]/\mathcal{B}[B^0\to Z_{c}(3900)K^0]=0.63\pm0.29$.   With  the same Wilson coefficients $a_1^{\prime}/a_{1}^{\prime *}$, the ratio  
$\mathcal{B}[B^+\to Z_{c}(3900)K^+]/\mathcal{B}[B^0\to Z_{c}(3900)K^0]$  becomes $0.98\pm0.39$, which shows no isospin breaking. As a result,  the isospin-breaking effect of the ratio  $\mathcal{B}[B^+\to Z_{c}(3900)K^+]/\mathcal{B}[B^0\to Z_{c}(3900)K^0]$ originates from the Wilson coefficients fitted to the experimental data.  

For $X_2(4013)$, in the particle basis, the ratio   $\mathcal{B}[B^{0}\to X_2(4013) K^{0}]/\mathcal{B}[B^{+}\to X_2(4013) K^{+}]$ is  estimated to be $0.75\pm0.16$.  With the same couplings $g_{n}^{\prime}/g_{c}^{\prime}$, the ratio becomes $0.92\pm0.20$.  With the same Wilson coefficients, the ratio becomes  $0.70\pm0.15$. Clearly, the isospin breaking of the neutral and charged components in its wave function is responsible for the large isospin breaking of this ratio.    


\bibliography{biblio.bib}

\end{document}